# Millimeter-wave Wireless LAN and its Extension toward 5G Heterogeneous Networks

**Kei SAKAGUCHI**[†], *member*, **Ehab Mahmoud MOHAMED**[†, ††], *non-member*, **Hideyuki KUSANO**[†], *member*,
**Makoto MIZUKAMI**[†††], *member*, **Shinichi MIYAMOTO**[†], *member*, **Roya REZAGAH**[†††], *non-member*,
**Koji TAKINAMI**, **Kazuaki TAKAHASHI**, *members*[††††], **Naganori SHIRAKATA**[††††], *non-member*,
**Hailan PENG**, **Toshiaki YAMAMOTO**, and **Shinobu NAMBA**, *members*[†††††]

**SUMMARY** Millimeter-wave (mmw) frequency bands, especially 60 GHz unlicensed band, are considered as a promising solution for gigabit short range wireless communication systems. IEEE standard 802.11ad, also known as WiGig, is standardized for the usage of the 60 GHz unlicensed band for wireless local area networks (WLANs). By using this mmw WLAN, multi-Gbps rate can be achieved to support bandwidth-intensive multimedia applications. Exhaustive search along with beamforming (BF) is usually used to overcome 60 GHz channel propagation loss and accomplish data transmissions in such mmw WLANs. Because of its short range transmission with a high susceptibility to path blocking, multiple number of mmw access points (APs) should be used to fully cover a typical target environment for future high capacity multi-Gbps WLANs. Therefore, coordination among mmw APs is highly needed to overcome packet collisions resulting from un-coordinated exhaustive search BF and to increase the total capacity of mmw WLANs. In this paper, we firstly give the current status of mmw WLANs with our developed WiGig AP prototype. Then, we highlight the great need for coordinated transmissions among mmw APs as a key enabler for future high capacity mmw WLANs. Two different types of coordinated mmw WLAN architecture are introduced. One is the distributed antenna type architecture to realize centralized coordination, while the other is an autonomous coordination with the assistance of legacy Wi-Fi signaling. Moreover, two heterogeneous network (HetNet) architectures are also introduced to efficiently extend the coordinated mmw WLANs to be used for future 5th Generation (5G) cellular networks.

## 1. Introduction

The exploding demand for the data traffic in cellular networks seems to be a persistent problem. According to [1], in 2014 almost half a billion (497 million) mobile devices and connections were added and the global mobile data traffic reached 2.5 exabytes per month, which shows 69 percent growth compared to 2013. This trend motivates a constant effort for implementing more efficient networks with higher capacity. Techniques such as massive MIMO [2], heterogeneous networks (HetNets) using small cells [3], and bandwidth expansion using mmw technology [4] are all introduced to address this issue. The main focus of this paper is on HetNets using coordinated mmw WLANs. Such an architecture not only takes the advantage of HetNet architecture but also benefits from huge available bandwidth in mmw band. This technique will be studied throughout this paper as a potential enabler for future 5G Networks.

Latest advances in mmw antennas and packaging technology [5] allow creating phased antenna arrays with limited number of elements. Next evolution in mmw technology is analog/digital hybrid beamforming (BF), like modular antenna arrays (MAA) [6] [7], consisting of a large number of sub-array modules. Each module has a built-in sub-array phase control and coarse beam steering capability, and fine BF is realized by digital signal processing. Such a high gain BF antenna has the potential to increase the range of mmw communication up to several hundreds of meters for line-of-sight (LOS) backhaul and several tens of meters for access [7] [8]. For communications in 60 GHz band, the ITU-R's recommended channelization comprises four consecutive channels, each 2.16 GHz wide, centered at 58.32, 60.48, 62.64 and 64.80 GHz respectively, which allows a single RF device to operate worldwide [9] - [17]. WiGig [18] and IEEE 802.11ad [19] are standards for the usage of 60 GHz band using single carrier (SC) and OFDM modulations with a maximum data rate of 6.7 Gbps [19]. Also, IEEE 802.11ad defines multi-band (2.4, 5 and 60 GHz) medium access control (MAC) protocol for backward compatibility with legacy IEEE 802.11 a, b, g, and n (Wi-Fi) standards [20] and for fast session transfer (FST) between them [19]. Antenna beamforming is usually used to combat 60 GHz channel propagation losses and accomplish data transmissions. Current IEEE802.11ad standard utilizes exhaustive search BF protocols on the MAC layer, using switched antenna array with a structured codebook, as a suitable BF mechanism for the 60 GHz band [18] [19] [21] [22]. As prototyping activities, in Oct. 2014, Qualcomm shows off WiGig with prototype tablets on Snapdragon 810 processors [23]. Also, Panasonic demonstrates a WiGig access point (AP) in early 2015, which will be described in more details in this paper.

All mmw architectures so far consider only point-to-point and point-to-multi-point applications [18] [19]. For future high capacity mmw WLANs, multi-point-to-multi-point transmissions, namely multiple 60GHz APs to cover multiple terminals, should be used to fully cover a typical target environment [24] - [30]. This architecture also provides the potential to enhance the robustness against shadowing and achieves higher spatial spectral efficiency by reducing the frequency reuse distance thanks to highly directional characteristics of 60 GHz frequency band. However, installing a numerous number of autonomously operated APs using CSMA/CA random access brings about several challenges related to the exhaustive search BF. For example, nearby APs may start the BF simultaneously, or an AP may perform the exhaustive search BF when its nearby APs are involved in data transmissions. Both cases cause a lot of packet collisions and degradation in the total system performance due to hidden nodes if the APs are autonomously operated. Therefore, a new system design is highly needed to coordinate mmw AP transmissions for future multi-Gbps WLANs. To the best of our knowledge, coordination of mmw APs for concurrent transmissions in random access scenario is never touched before in the literature, while the authors in [31] proposed a coordinated beamforming assuming perfectly synchronous network. In this paper, two types of coordinated mmw WLAN architecture are presented, which we proposed in [26]–[30]. The MAC protocols required to operate the coordinated mmw WLANs are also given. The first mmw WLAN architecture considers the mmw APs as remote radio heads (RRHs) connected to a WiGig AP base band unit (BBU), which coordinates the BF and data transmissions operations among the installed APs in a full centralized manner [29] [30]. The second architecture is a sub-cloud mmw WLAN, which utilizes the wide-coverage 5GHz (Wi-Fi) signal to autonomously coordinate the transmissions among dual-band (5, 60 GHz) mmw APs using a novel dual-band MAC protocol based on statistical learning [26]- [28].

---

†     The author is with Osaka University, Osaka, Japan.
††    The author is with Aswan University, Aswan, Egypt.
†††   The author is with Tokyo Institute of Technology, Tokyo Japan.
††††   The author is with Panasonic Corporation, Japan.
††††† The author is with KDDI R&D Laboratories, Inc., Japan.

In this paper, we also study the efficient interworking between mmw WLAN and cellular networks as a 5G enabler. Coordinated mmw WLANs can provide an extraordinary high data rate which can be efficiently distributed in hotspot areas. However, due to high propagation loss and penetration loss, the coverage area of mmw WLANs will be in the range of several tens of meters [7] [8]. This issue makes it much more difficult for a user equipment (UE) to discover mmw small WLANs and handover to them. Here, the general concepts of HetNets [32]-[34] and Control-plane/User-plane splitting (C/U splitting) [35] are exploited to facilitate the LTE/mmw WLAN internetworking. In this paper, three HetNet architectures, including 3rd generation partnership project (3GPP) standard [36], are introduced as candidates of LTE/mmw WLANs internetworking. In contrast to the loosely coupled LTE/WLAN internetworking in 3GPP using access network discovery and selection function (ANDSF) [37], a tightly coupled LTE/WLAN internetworking, proposed by the authors in [38], is presented. In addition, a multi-layer LTE/WLAN/WiGig internetworking based on the concept of inter-RAN exchange (IRX) with enhanced ANDSF (E-ANDSF), proposed by the authors in [39], is also given as a candidate for LTE/mmw WLAN internetworking. Pros and cons of each architecture candidate are explained in details in this paper.

The rest of this paper is organized as follows. Section 2 presents an overview of the current WiGig/IEEE802.11ad based WLAN and also describes a possible extension toward multiple mmw AP coordination. Section 3 provides the sub-cloud mmw WLAN where Wi-Fi and WiGig coordinate tightly via dual-band MAC protocol based on statistical learning. Further extension toward 5G HetNet using mmw WLANs are given in Sect. 4. Finally, Sect. 5 concludes this paper. Table 1 summarizes important abbreviations used in this paper.

**Table 1** Abbreviations.

| Acronym | Description |
|---|---|
| 5G | $5^{th}$ generation |
| A-BFT | association BF training |
| ACK | acknowledgement |
| ANDSF | access networks discovery and selection function |
| APC | AP controller |
| API | AP information |
| ATI | announcement transmission interval |
| BBU | base band unit |
| BF | beamforming |
| BHI | beacon header interval |
| BID | best beam identification |
| BLI | blocking information |
| BRP | beam refinement protocol |
| BS | base station |
| BT | beam tracking |
| BTI | beacon transmission interval |
| C/U | control-plane/user-plane |
| CAPEX | capital expenditure |
| CBAP | contention-based access period |
| CEF | channel estimation field |
| CLI | candidate link information |
| CMOS | complementary metal-oxide-semiconductor |
| CSMA/CA | carrier sense multiple access/collision avoidance |
| CTS | clear to send |
| DB | database |
| DBPSK | differential BPSK |
| DTI | data transfer interval |
| E-ANDSF | enhanced ANDSF |
| EIRP | equivalent isotropic radiated power |
| EPC | evolved packet core |
| ePDG | evolved packet data gateway |
| E-UTRAN | evolved universal terrestrial RAN |
| FBK | feedback |
| FST | fast session transfer |
| HDMI | high definition multimedia interface |
| HetNet | heterogeneous network |
| IC | integrated circuit |
| IRX | inter-RAN exchange |
| LDO | low drop-out regulator |
| LOS | line-of-sight |
| LP | learning period |
| MAA | modular antenna array |
| MAC | media access control |
| MCS | modulation and coding scheme |
| mmw | millimeter-wave |
| MNO | mobile network operator |
| NAV | network allocation vector |
| NF | noise figure |
| OFDM | orthogonal frequency division multiplexing |
| OPEX | operating expense |
| PAPR | peak to average power ratio |
| PDN | packet data network |
| PHY | Physical |
| PMU | power management unit |
| PS | phase shifter |
| RAN | radio access network |
| RRH | remote radio head |
| RSS | radio signal strength |
| RSSI | RSS indicator |
| RTS | request to send |
| SC | single carrier |
| SIFS | short inter frame space |
| SLS | sector-level sweep |
| SP | service period |

| STA | Station |
|-----|---------|
| STF | short training field |
| UE | user equipment |
| USB | universal serial bus |
| WLAN | wireless local area network |

## 2. WiGig/IEEE 802.11ad based WLAN and its extension

Since the WiGig/IEEE 802.11ad enables multi-Gbps throughput, it is expected to spread as a high-speed wireless alternative to existing wired standards such as HDMI or USB 3.0 as illustrated in Fig. 1(a). Even though the first target will be to replace the wired high-speed interface, the 60 GHz wireless is also capable of providing high-speed multiuser access that can be deployed in dense small cell networks as in train stations shown in Fig. 1(b). However, since the WiGig/IEEE 802.11ad are based on the CSMA/CA protocol, the throughput is significantly degraded under multiuser environment, necessitating the extension of the existing protocol in order to maintain high throughput for multiuser access.

This section starts with a brief overview of the PHY and MAC specifications in WiGig/IEEE 802.11ad, and then explains future extension activities for the 60 GHz WLAN. The section also describes hardware implementations including a mobile terminal and an AP prototype.

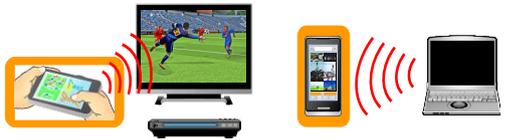

(a) Low latency video streaming and fast file transfer.

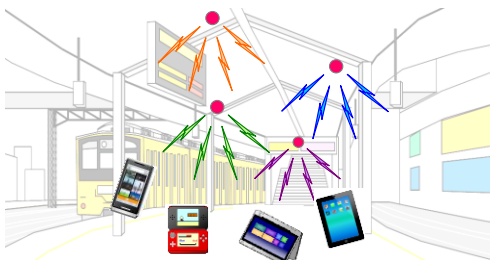

(b) High-speed multiuser wireless access
**Fig. 1** 60 GHz use case examples.

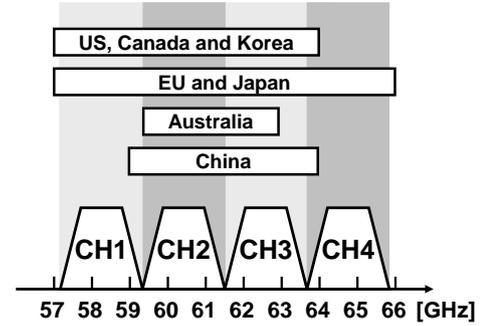

**Fig. 2** Frequency allocations in 60GHz band.

**Table 2** 60 GHz regulations among different regions.

| | | EIRP | Output power ($P_o$) and antenna gain ($G_a$) |
|---|---|---|---|
| Japan | | Not specified | $P_o \leq 10$ dBm (10 mW) <br> $G_a \leq 47$dBi |
| US | Outdoor | 1. $G_a < 51$ dBi <br> EIRP = 82 dBm − 2 (51- $G_a$) <br> 2. $G_a \geq 51$ dBi <br> EIRP = 82dBm | |
| | Indoor | 40 dBm | $P_o \leq 27$ dBm (0.5 W)* |
| Korea | | 43 dBm or 57 dBm (Fixed point-to-point) | 1. Directional ant. <br> $P_o \leq 27$ dBm (0.5 W) <br> 2. Omni directional ant. <br> $P_o \leq 20$ dBm (0.1 W) |
| EU | | 40 dBm* | Not specified |

*Specified in spectrum density for a narrow band signal

**Table 3** Example of modulation and coding schemes.

| MCS Index | Modulation | Code Rate | PHY Payload Rate [Mbps] |
|-----------|------------|-----------|-------------------------|
| 0 | DBPSK (SC) | ½ | 27.5 |
| 1 | π/2-BPSK (SC) | ½ | 385 |
| 4 | π/2-BPSK (SC) | ¾ | 1155 |
| 5 | π/2-BPSK (SC) | 13/16 | 1251.25 |
| 9 | π/2-QPSK (SC) | 13/16 | 2502.5 |
| 12 | π/2-16QAM (SC) | ¾ | 4620 |
| 21 | 16-QAM (OFDM) | 13/16 | 4504.5 |
| 24 | 64-QAM (OFDM) | 13/16 | 6756.75 |

Note: MCS 0 and MCS 1 employ spreading factor of 32 with π/2 rotation and spreading factor of 2 respectively.

### 2.1 PHY/MAC standard in WiGig/IEEE 802.11ad

The most recent global frequency allocation at 60 GHz is illustrated in Fig. 2. Frequency bands around 60 GHz are available worldwide. Table 2 summarizes 60 GHz regulations among different regions. As can be seen in the table, the maximum antenna power allowed in Japan is 10 dBm, which is lower than other regions. It is currently under discussion for relaxing the regulation from the viewpoint of global uniformity[1]. The output power and the antenna gain vary depending on the process technology and antenna configurations. Even though GaAs or SiGe processes can deliver a high output power up to around 20 dBm, the low-cost CMOS solution is preferred for most of consumer applications. In general, the output power in today's advanced CMOS technology is limited to much lower than 10 dBm due to transistors' reliability issue. To extend the communication distance, the beamforming technology which consists of an array antenna with beam steering capability is widely used to obtain higher antenna gain, at the cost of increased power dissipation. A recent publication [40] employs 16 Tx and 16 Rx array structure, demonstrating 4.6 Gbps throughput at 10 m communication distance while dissipating 960 mW in Rx and 1190 mW in Tx excluding the power consumption of the baseband signal processing.

---

[1] The new regulation is expected to be effective in 2015 where the maximum output power up to 24 dBm (250 mW) will be allowed.

In WiGig/IEEE 802.11ad, both SC modulation and OFDM modulation have been adopted, considering various use case scenarios. In general, SC modulation is suitable for reducing power consumption due to its low peak-to-average power ratio (PAPR), whereas OFDM modulation offers better multipath tolerance. Table 3 lists an example of the modulation and coding schemes (MCSs) where MCS 0 to MCS 4 are mandatory. MCS 0 is exclusively used to transmit control channel messages employing differential BSPK (DBPSK) modulation with code spreading to ensure better robustness. The SC modulation supports up to 4.62 Gbps (π/2-16QAM) whereas the OFDM modulation achieves 6.75 Gbps (64-QAM) maximum PHY throughput rate.

Figure 3 shows the packet structure examples. The packets consist of a short training field (STF) and a channel estimation field (CEF), followed by a header, data and subfields. The subfields can be used for fine beamforming training.

To achieve multi-Gbps throughput, the MAC layer requires many modifications from the one in the traditional IEEE 802.11. For example, a minimum interval time between transmission packets, called short inter frame space (SIFS), is shortened from 16 μs in the IEEE 802.11a/g to 3 μs in the WiGig/IEEE 802.11ad. Many timing parameters of random backoff in CSMA/CA are also shortened. On the other hand, the maximum payload size is extended from 2304 octets to 7920 octets. These modifications reduce packet overhead and improve transmission efficiency.

As illustrated in Fig. 4, channel access occurs during beacon intervals and is coordinated using a schedule. The beacon interval comprises a beacon header interval (BHI) and a data transfer interval (DTI). The BHI consists of a beacon transmission interval (BTI), and optional association beamforming training (A-BFT) or announcement transmission intervals (ATI). The DTI can include one or more scheduled service periods (SPs) and contention-based access periods (CBAPs). In a typical setup, the beacon interval is set to be around 100 ms [19]. Since the propagation characteristic of 60 GHz communications is highly directional, the Tx and Rx antenna beam pattern need to be aligned in the right direction to obtain sufficient gain in the BHI.

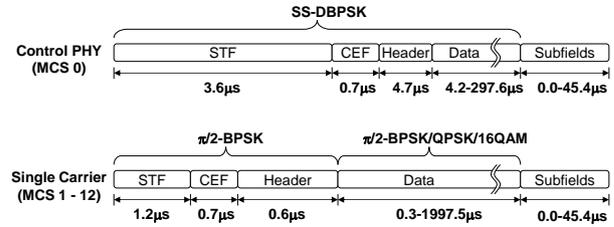

**Fig. 3** Packet structure examples.

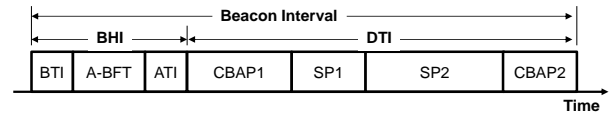

**Fig. 4** Example of access periods within beacon interval.

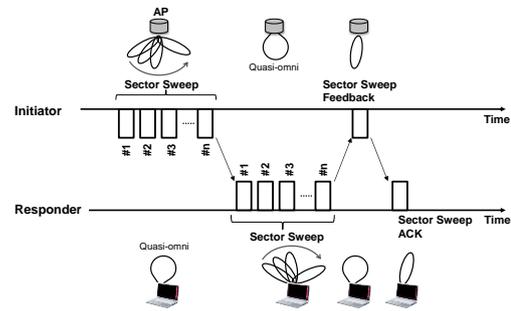

**Fig. 5** Example of beamforming training.

The WiGig/IEEE 802.11ad employs bidirectional sequence of beamforming training frame transmission that uses sector sweep and provides the necessary signaling. The beamforming protocol involves sector-level sweep (SLS), beam refinement protocol (BRP) and beam tracking (BT). Figure 5 illustrates an example of the SLS, which is the most basic type of Tx beam direction training. During the BTI, the initiator sends sector sweep frames as training signals while the responder measures the quality of the received frames using a quasi-omni beam pattern which is the widest beam width attainable. This can be realized by simply using one antenna element which achieves wider beam width compared to an array antenna. Then, the responder feedbacks the best sector ID while changing the responder's beam patterns. Similar process can also be applied to align Rx beam direction if needed. The BRP phase after the SLS is to enable iterative refinement of its antenna configuration (or antenna weight vector) of both the transmitter and the receiver. Both SLS and BRP use their own special frames for BF training. In contrast, the BT is implemented by adding training fields at the end of a data packet, which is used to adjust the antenna configuration during data transmission.

## 2.2 Next generation of IEEE802.11ad

The use of high frequencies comes with both advantages and disadvantages. Large path loss, as well as high attenuation due to obstacles such as a human body, limits 60

GHz applications to those suitable for short-range point-to-point communication. On the other hand, higher frequencies lead to smaller sizes of RF components including antennas, enabling compact realization of an array structure which offers larger antenna gain with high directivity as mentioned earlier. These properties unique to 60 GHz help reducing interference among terminals, and offering opportunities for realizing high speed point-to-multipoint or multipoint-to-multipoint connections as a future extension.

Since communication distance at 60 GHz is limited, each AP forms a small cell with 3 to 10 m radius and the multiple APs will be used for sufficient area coverage. In such a 60 GHz wireless network, the AP cooperative beamforming will be highly required to enhance system throughput as well as to alleviate link disconnection due to shadowing. These enhancements require efficient beamforming protocol, resource management, etc., as it will be explained in more details throughout this paper.

Standardization for the next generation 60 GHz wireless has already started in IEEE 802.11WG and it is under discussion in the Next Generation 60 GHz Study Group (NG60SG)[2]. The study group aims to achieve maximum throughput of 20 Gbps while maintaining backward compatibility and coexistence with legacy WiGig/IEEE 802.11ad, targeting various use cases such as multiuser mass data distribution, wireless backhaul, ultra-short-range communication, etc. NG60SG also aims to improve the spatial reuse of simultaneous nearby transmissions to increase the aggregated system throughput. It is expected to employ higher-modulation, channel-bonding, and MIMO technology (polarization-based, SVD-based, or hybrid-beamforming-based) for throughput enhancement as well as multiuser MIMO and interference suppression for high throughput multiuser operation. The introduction of the new standard is targeting around 2020.

## 2.3 Prototype Hardware of WiGig/IEEE802.11ad WLAN

This sub-section presents hardware prototypes based on WiGig/IEEE 802.11ad WLAN.

### 2.3.1 Prototype of the Mobile Terminal

First example is the low power solution targeting mobile applications. Figure 6 shows a block diagram and a die photo of the RFIC [40]. It employs direct conversion architecture that is advantageous to reduce power consumption. Power management unit (PMU) integrates low drop-out regulators (LDOs) that provide regulated 1.25 V/1.4 V DC voltages from 1.8 V power supply to improve external noise tolerance. Since the WiGig/11ad employs wide modulation bandwidth of 2.16 GHz per channel, the transceiver suffers from large frequency dependent distortion caused by gain variations of analog circuits as well as multipath environment. To minimize performance degradations, the transceiver employs sophisticated digital calibration schemes, such as built-in Tx in-band calibration and an Rx frequency domain equalizer (FDE). These techniques relax the requirement for high speed analog circuits, leading to less power consumption with minimum hardware overhead [41].

The RFIC is fabricated in a 90 nm CMOS process with 3.1 x 3.75 mm chip size. The RFIC consumes 347 mW in the Tx mode and 274 mW in the Rx mode. As shown in Fig. 7, the RFIC is integrated in the cavity structure antenna module with 10 x 10 mm small size. Each Tx/Rx antenna is printed on the surface of the antenna module which consists of four patch elements, providing 6.5 dBi gain with 50 degree beam width. The output power is set to be 2 dBm, which corresponds to 6 dB back-off from the saturated output power of 8 dBm, providing 8.5 dBm EIRP. The relatively low output power is chosen to avoid transistor performance degradation due to voltage stress in order to guarantee 10-year lifetime. The measured Rx NF is 7.1 dB. The antenna module is mounted on the 7x4 cm evaluation board with a baseband IC including PHY and MAC layers. The board is connected to a laptop PC through USB 3.0 interface, achieving over 1.7 Gbps maximum MAC (2.5Gbps PHY) throughput at MCS 9.

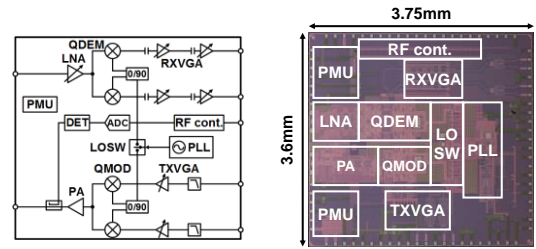

**Fig. 6** Block diagram (left) and die photo (right) of RFIC.

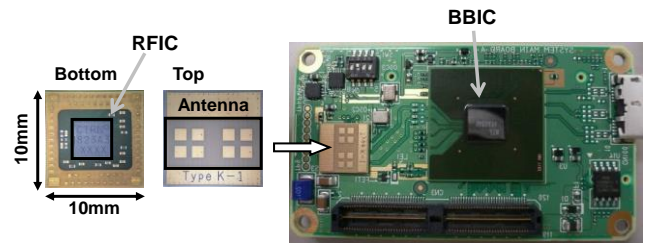

**Fig. 7** Antenna module (left) and an evaluation board with USB interface (right).

### 2.3.2 Prototype of the Access Point

The second example is the prototype hardware for APs in which wider communication coverage is required. At the mmw frequency, it is possible to implement BF by using a phased array antenna in a small form factor. The required phase shift for each antenna element can be introduced in either RF, local or baseband. In order to achieve small chip area with reduced power consumption, we adopt RF phase shifter (PS) approach where only the low noise amplifier (LNA) and the power amplifier (PA) need to be duplicated. The main challenge of this approach is the implementation of an RF PS that operates at 60 GHz band. Figure 8 (a) and (b) show the PS circuit and the block diagram of the RFIC [42]. The RFIC employs RF PSs for 4Tx/4Rx signal paths. The PS utilizes a quadrature hybrid with a digitally controlled vector combiner. The transformer-based magnetic coupling hybrid offers a small footprint with excellent amplitude/phase accuracy. The vector combiner is composed of 5-bit current-controlled RF amplifiers, providing 32 x 32 variable outputs. By flipping the polarity of differential IQ signals, it covers one of the four quadrants in the IQ signal space, providing 5° phase resolution over a 360° variable range.

The RFIC is fabricated in a 40 nm CMOS process and it is integrated in the miniaturized antenna module shown in Figure 8 (c). The antenna module is 11 x 12 mm and it achieves 7 dBi maximum antenna gain, delivering the total Tx power of 10 dBm EIRP. Even though the four Tx path configuration allows to achieve higher output power, the output power is set to 3 dBm in the prototype to mitigate thermal constraints. Simple codebook based beamforming offers about 120° beam steering range with 7 steps in the azimuth direction.

Figure 9 shows measured MAC throughput over the air. In this measurement, two evaluation boards are fixed to the moving positioners that can control both communication distance and their direction. As shown in Fig. 9 (a), the measured MAC throughput achieves 1.7 Gbps (MCS 9) up to 3 m and 1.0 Gbps (MCS 5) up to 5 m in the maximum antenna gain direction. The area coverage is also measured by rotating one of the boards in the azimuth direction. Beam direction is controlled based on the WiGig/IEEE 802.11ad BF protocol. As shown in Fig. 9(b), it achieves 1.7 Gbps MAC throughput at more than 2.5 m communication distance while covering +/- 60 deg angle range.

By using the fabricated antenna modules, a prototype of the AP is designed. As illustrated in Fig. 10, the combination of the three antenna modules, covers 360° around the AP, while supporting up to three concurrent communication links with spatial sharing. It may be required to extend communication distance to cover larger area by one AP. This is realized by increasing the number of RF signal paths to deliver higher output power as well as to enhance antenna gain, which is a subject of future works.

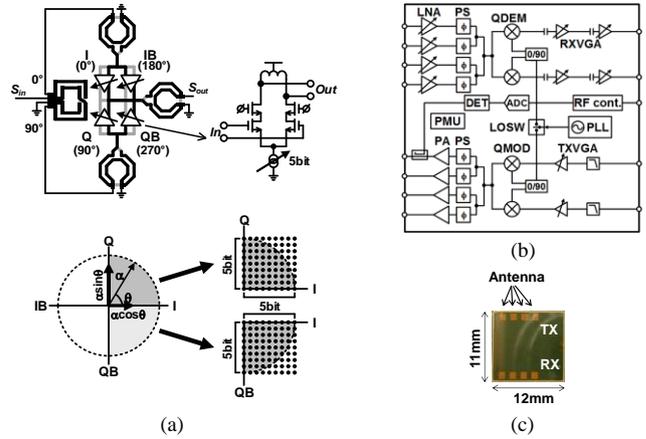

**Fig. 8** (a) Phase shifter circuits, (b) block diagram of RFIC and (c) picture of antenna module.

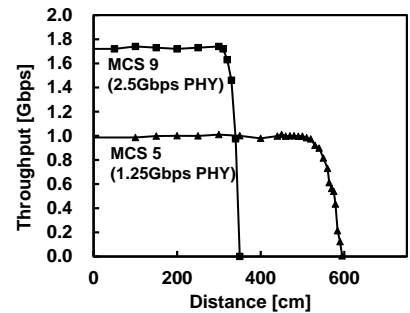

(a) Throughput vs. Communication distance

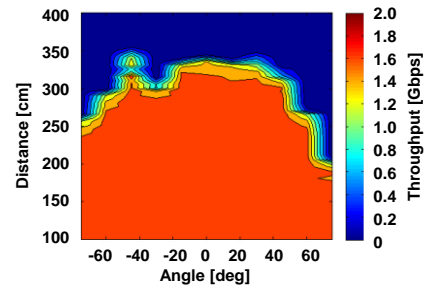

(b) Area coverage (with MCS 9)

**Fig. 9** Measured MAC throughput over the air.

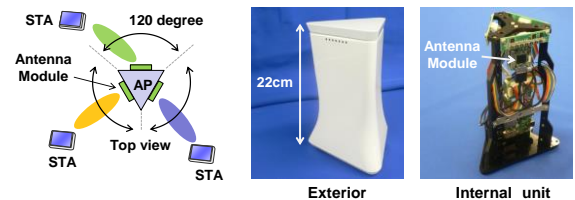

**Fig. 10** AP structure with three antenna modules (left) and the picture of the AP prototype (right).

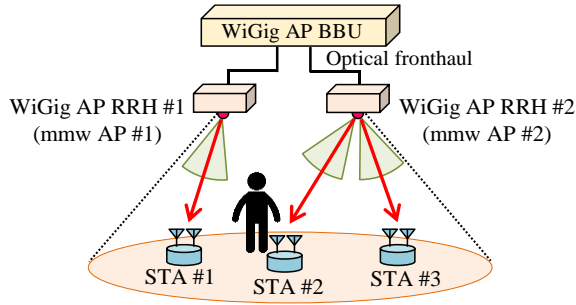

**Fig. 11** Distributed RRH architecture for mmw APs coordination.

## 2.4 Extension to multi-mmw AP coordination by using WiGig RRH architecture

Multiple mmw APs are required to be installed in the future mmw WLANs to fully cover a target environment and overcome shadowing problem. However, installing numerous number of mmw APs imposes several coordination challenges due to the exhaustive BF training with CSMA/CA random access. This subsection presents an example of multi-mmw AP coordination for future mmw WLAN, which we proposed in [29] [30]. The key idea of this mmw WLAN is to distribute WiGig AP RRHs in the target environment and control them in a coordinated manner via WiGig AP BBU. By employing RRH architecture, synchronously coordinated BF among multiple mmw APs can be achieved by simply extending the current MAC protocol shown in Sec. 2.1. From now on, we call WiGig AP RRH as mmw AP and WiGig AP BBU as BBU for simpler notation. Figures 11 and 12 illustrate an example of the coordinated mmw WLAN and its MAC protocol, respectively. In this example, two mmw APs are connected to a BBU through optical fronthaul and three stations (STAs) coexist in the same coverage area.

### 2.4.1 Coordinated Beamforming Training

In traditional MAC protocol defined in IEEE 802.11ad and given in Sec. 2.1, each client STA associates with the AP achieving the highest received signal strength (RSS) value. In contrast, in the MAC protocol shown in Fig. 12, during the BHI, each STA tries to associate with all mmw APs that have RSS values higher than a specified minimum threshold, and the best transmission sectors for all links between APs and associated STAs are determined. At first, to initiate the BTI, the BBU sends a trigger frame for starting sector sweep to the mmw APs. All mmw APs and STAs keep silent until receiving the trigger frame. After receiving the trigger frame, the representative mmw AP pre-determined by the BBU sends sector sweep frames, and all the other APs and STAs measure the RSS values using a quasi-omni beam pattern. In the example shown in Fig. 12, AP #1 is a representative AP and it sends sector sweep frames. Since all other APs and STAs (AP #2 and STA #1-#3) still keep silent during the transmission of sector sweep frames, these APs and STAs can measure RSSI from the AP #1 without any interference. After the transmission of sector sweep frames from AP #1, AP #2 starts to send sector sweep frames and RSS values from AP #2 is measured by AP #1 and all STAs. When transmission of sector sweep frames from all mmw APs is completed, each STA initiates to send its sector sweep frames according to the order specified in sector sweep frames sent from mmw APs. Sector sweep frames sent from each STA is also utilized to notify the best sector ID for each AP. After the transmission of sector sweep frames from STA, each AP replies sector sweep feedback frame to notify the best sector to the corresponding STA. After transmission of sector sweep and sector sweep feedback frames from all STAs, mmw APs send RSS indicator feedback (RSSI feedback) signal to notify the best sectors and RSS values of all links to the BBU. By using RSSI feedback signal, BBU calculates signal to interference pulse noise ratio (SINR) for all combinations of best sectors as shown in Fig. 13. Based on the calculated SINRs, the BBU selects AP-STA combinations that satisfy the minimum required SINR as possible candidates used for the transmission of RTS, CTS and DATA frames during DTI. Selected combinations are notified to mmw APs by sending candidate link information (CLI) signal.

The explained coordinated sector sweep incurs significant signaling overhead especially at BHI to avoid interference between BF trainings. For example in the case of $N$ APs and $N$ STAs, $N$ times period for sector sweeps is needed compared to the single AP and single STA transmission, while on the other hand at most $N$ simultaneous transmissions are available. With rough estimation, the total throughput of $N$ simultaneous transmission $R_N$ can be estimated as:

$$R_N = (1 - N \times \alpha_1) \times N \times R_1, \qquad (1)$$

where $R_1$ and $\alpha_1$ are average throughput and overhead ratio in the case of single AP and single STA transmission, and it is assumed that $N \times \alpha_1 < 1$. From the equation, it is obvious that the scheme is effective when $\alpha_1$ is small. This implies that, in the proposed protocol, DTI should be long enough compared to the BHI, however it sacrifices tolerance against the mobility. Another coordination technology that is also able to reduce the overhead will be discussed in Sect. 3 where location based beamforming is applied by using Wi-Fi fingerprint.

### 2.4.2 Coordinated DATA Frame Transmission

In DTI, prior to transmission of DATA frames, to reserve the radio resource and detect the disconnection of the links due to human shadowing, all mmw APs and associated STAs exchange RTS/CTS frames. For example, as shown in Fig. 12, when three STAs are associated with two mmw APs, the exchanges of six RTS/CTS frames (RTS #1-#6, CTS #1-#6) are required. To shorten the duration required for exchanging the all RTS/CTS frames, the BBU selects the combination of links which minimize the required duration among all possible combinations. When the mmw AP cannot receive CTS frame from the associated STA, the mmw AP decides that the link is disconnected and notifies it to the BBU by sending a blocking information (BLI) signal. After receiving BLI signal, the BBU determines the link (i.e. mmw AP) employed for the transmission of DATA frame to each associated STAs among the remaining combinations of the links. Several types of scheduling algorithms can be introduced to determine the links employed for the transmission of DATA frames. Determined links are notified to the mmw APs and the associated STAs by sending assigned mmw AP information (API) signal. After the transmission of API signal, each mmw AP transmits its DATA frame to the corresponding associated STAs. Finally, if the DATA frame is correctly received, the associated STA replies an ACK frame to the corresponding mmw AP. As shown in Fig. 12, in the proposed MAC layer protocol, to reduce the frame transmission error as much as possible, STAs transmit their ACK frames in a time division manner, and the time to send ACK frames are specified in API signal. To increase the MAC efficiency of the proposed protocol, simultaneous transmission of ACK frames from multiple STAs can be introduced thanks to the space division multiple access based on beamforming. When the mmw AP does not receive the ACK frame, the mmw AP decides that the link employed for the transmission of DATA frame is blocked and notifies it to the BBU. Then, the BBU re-coordinates the allocation of link employed for the re-transmission of DATA frame. Since mmw AP coordination enables WLAN to dynamically select the mmw AP employed for the transmission of DATA frames to the associated STA in accordance with the disconnection of the links and received SINR, it avoids the degradation of throughput caused by human shadowing and collision due to hidden STA problem.

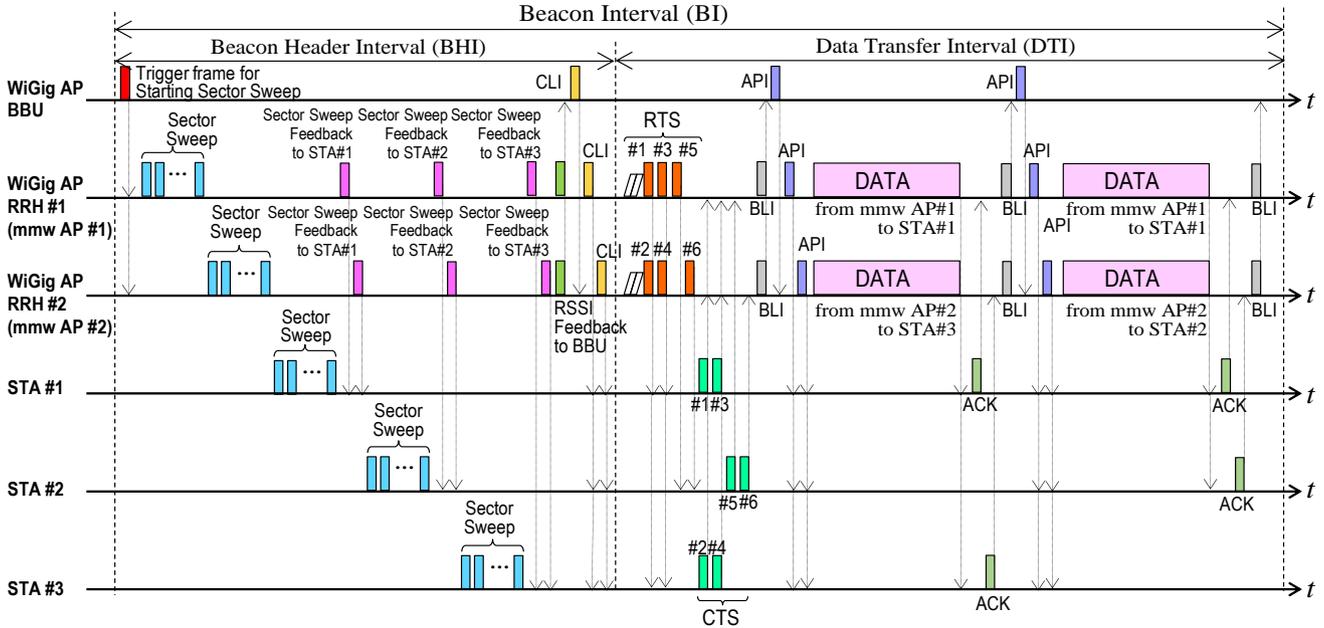

**Fig. 12** MAC layer protocol designed for WLAN with multi-mmw AP coordination.

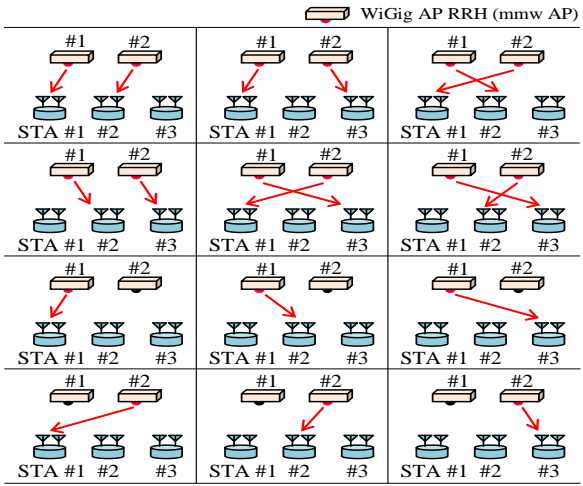

**Fig. 13** Combinations of the links between mmw APs and STAs.

## 3. Wi-Fi/WiGig Coordination for Sub-Cloud WLAN

Since the Wi-Fi alliance has integrated the WiGig alliance, the future WLAN chipsets must support multi-band such as 2.4, 5, and 60GHz. In IEEE 802.11ad, FST [19] is introduced for enabling STAs to transfer the ongoing data transmission from one band to another band based on the RSS of the different bands. IEEE 802.11ad also introduces single MAC address (virtual MAC address) for both Wi-Fi and WiGig to facilitate the FST operation. This FST works effectively only in the case of single AP where handover is occurred only between the homothetic coverage of Wi-Fi and WiGig. In the case of multiple APs, we should consider the handover between APs as well as avoiding interference between the beams from different APs. In this section, we introduce more comprehensive coordination between Wi-Fi and WiGig to realize future high capacity multi-Gbps WLANs by using sub-cloud architecture.

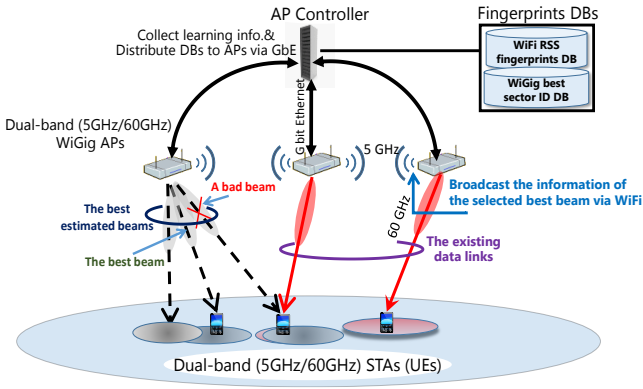

Fig. 14 Sub-cloud WLAN system architecture.

## 3.1 Sub-cloud WLAN using Wi-Fi/WiGig Coordination

Figure 14 shows system architecture of the sub-cloud WLAN using Wi-Fi/WiGig coordination [26]. In Fig. 14, multiple dual-band (5 and 60 GHz) APs are connected to an AP controller (APC) via gigabit Ethernet links. This sub-cloud WLAN will be installed in a target environment to cover it using multiple dual-band APs. In this architecture, the APC works as a central coordinator of the sub-cloud to assist radio resource management of multiple APs by keeping the random access protocol in Wi-Fi/WiGig networks. Moreover, the APC acts as an interface between the sub-cloud and cellular networks that will be explained in Sec. 4. To facilitate the coordination process among dual-band APs, several technologies are used in the proposed WLAN.

The first technology is an extension of the current FST to use Wi-Fi RSS (fingerprint) for location (beam) management of the UEs (APs). If more than two APs are installed in the target environment, we can roughly identify the location of a UE by reading Wi-Fi RSS values of multiple APs that is called fingerprint in this paper. Since the best AP to be associated and the best beam to be selected are location (fingerprint) dependent, the exhaustive search BF training of multiple APs is not needed anymore if we have a database (DB) to make links between fingerprints and best beam IDs. For that purpose, offline statistical learning [28] is introduced in this paper. Moreover, the bad beams from other APs to make destructive interference to the current AP-UE link can also be estimated by using the fingerprint and DB, that bad beams should be eliminated from the BF training in other APs. Thus, efficient coordinated beamforming among multiple mmw APs without exhaustive search BF training can be achieved by Wi-Fi/WiGig coordination in the sub-cloud WLAN.

Another technology is the introduction of C/U splitting

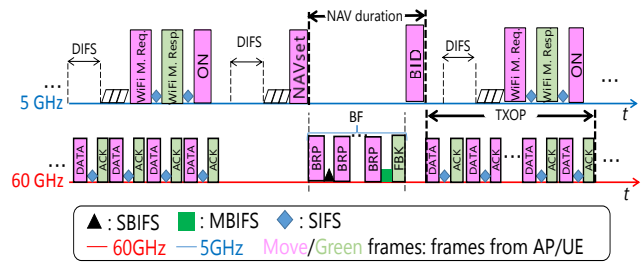

Fig. 15 Dual-band MAC protocol for coordinated WLAN.

[35] over dual-band MAC protocol in the sub-cloud WLAN, where control frames to be shared with multiple APs are transmitted by Wi-Fi while the high speed data frames are transmitted by WiGig concurrently from multiple mmw APs. The control frames to be transmitted by Wi-Fi includes the frames needed for Wi-Fi fingerprint measurement, a Network Allocation Vector set (NAVset) frame to avoid collision during the beam refinement, and a best beam identification (BID) frame to inform the selected beam to be used. Since the coverage of Wi-Fi is much larger than that of WiGig, the Wi-Fi frames are used to manage location of UEs, to avoid packet collision between mmw APs, and to inform beam ID information selected autonomously in each mmw AP. Such a tight coordination among Wi-Fi and WiGig enables concurrent transmission of multiple data frames from different mmw APs, and also realizes seamless handover among mmw APs as well as between Wi-Fi and WiGig. It is noted that most of the data traffic is offloaded to WiGig, so that the additional control traffic on Wi-Fi is almost negligible and collisions on control frame hardly happen.

## 3.2 Dual-band MAC Protocol for Coordinated mmw Transmissions

Figure 15 shows the dual-band MAC protocol, which we proposed to coordinate multiple mmw transmissions in random access scenarios [27]. In this protocol, if a data frame is generated for a specific UE, the random access process starts using Wi-Fi interface instead of mmw. In the beginning, one of unused APs senses the 5GHz band using carrier sense routine. If the medium is free, it starts the backoff counter. If the counter reaches zero, it starts to send a Wi-Fi measurement request (Wi-Fi M. Req.) frame to intended UE. Then, the UE broadcasts a Wi-Fi measurement response (Wi-Fi M. Resp.) frame for measuring its current Wi-Fi fingerprint. Based on the current Wi-Fi fingerprint and offline fingerprint DBs, the APC associates the UE to an appropriate unused AP. Then, it sends a switch ON frame to the UE from the unused AP using its Wi-Fi interface to turn ON the mmw interface if it's in a sleep mode. The APC also estimates a group of best beams for the selected AP-UE link and groups of bad beams that produce interference to this link from other APs. Then, the APC sends this best and bad beam information to the coordinated mmw APs. After this preparation process using Wi-Fi interface, the selected AP starts the random access process again in the 5GHz band. If the backoff counter reaches zero, a NAVset frame is sent from the AP using its Wi-Fi interface to prevent any other AP from doing BF refinement in 60GHz band until it finishes. The NAVset frame contains the estimated time that the AP will take until it finishes BF refinement. The BF refinement selects the best beam from the group of best beams by using high speed BRP frames. At the end of BF refinement process, UE sends a feedback (FBK) frame to the AP using the 60 GHz band to inform the ID of the highest link quality beam (the best beam) and its received power. Consequently, the AP broadcasts a best beam Identification (BID) frame, which contains the estimated best beam ID using Wi-Fi signaling. By broadcasting the BID frame, other APs will consider the selected link information when they accurately estimate their bad beams before conducting BF refinement. After broadcasting the BID, the AP starts to send data frames to the UE using the coordinated best beam in 60 GHz band.

### 3.3 Best Beams Estimation and Bad Beams Elimination Using Statistical Learning

To efficiently coordinate the BF process among the different mmw APs and reduce packet collisions during BF, the concept of best and bad beams are introduced. The best beams are estimated based on the current UE Wi-Fi fingerprint (location) by using statistically learned DBs to speed up the BF training by removing the exhaustive search mmw BF in the SLS phase. Among these estimated best beams, beam directions that degrade the data rate of the existing link (bad beams) are eliminated from the BF refinement. The candidates of bad beams can be estimated by evaluating the overlap of best beams of different mmw APs in the best beam DB. In the online phase, more accurate bad beams are selected by using the BID frame sent from the existing link.

#### 3.3.1 Offline Statistical Learning Phase

##### 3.3.1.1 Collecting Fingerprints Databases (DBs)

The first step in the offline statistical learning phase is to construct the Wi-Fi and mmw radio maps for target environments. Constructing the radio maps can be effectively done by collecting the average Wi-Fi RSS readings (fingerprints) and mmw APs best sectors IDs at arbitrary learning points (LPs) in the target environment. Therefore, three databases are constructed as radio maps, i.e. the Wi-Fi fingerprint DB $\mathbf{\Psi}$, the best sector ID DB $\mathbf{\Phi}$, and the offline received power DB $\mathbf{P_{OFF}}$, which are defined as:

$$\mathbf{\Psi} = \begin{pmatrix} \psi_{11} & \cdots & \psi_{L1} \\ \vdots & \ddots & \vdots \\ \psi_{1N} & \cdots & \psi_{LN} \end{pmatrix}, \mathbf{\Phi} = \begin{pmatrix} \phi_{11} & \cdots & \phi_{L1} \\ \vdots & \ddots & \vdots \\ \phi_{1N} & \cdots & \phi_{LN} \end{pmatrix},$$

$$\mathbf{P_{OFF}} = \begin{pmatrix} p_{11}^{\phi_{11}} & \cdots & p_{L1}^{\phi_{L1}} \\ \vdots & \ddots & \vdots \\ p_{1N}^{\phi_{1N}} & \cdots & p_{LN}^{\phi_{LN}} \end{pmatrix}, \tag{2}$$

where $\psi_{ln}$ is the Wi-Fi fingerprint at AP $n$ from the UE located at LP $l$. $L$ is the total number of LPs, and $N$ is the total number of APs. $\phi_{ln}$ is the best sector ID at LP $l$, which corresponds to the sector ID of AP $n$ that maximizes the received power of UE located at LP $l$. $\phi_{ln}$ can be calculated as:

$$\phi_{ln} = d_n^* = \arg\max_{d_n}\big(P_{ln}(d_n)\big), \ 1 \le d_n \le D_n, \tag{3}$$

where $d_n$ indicates the sector ID of AP $n$, $D_n$ is the total number of sector IDs, and $\phi_{ln} = d_n^*$ is the best sector ID at LP $l$ from AP $n$ that maximizes the received power $P_{ln}(d_n)$. A *null* sector ID in the $\mathbf{\Phi}$ matrix, i.e. $\phi_{ln} = null$, means that AP $n$ cannot cover LP $l$. The *null* values are used by the APC for the Wi-Fi/mmw association/re-association decisions. $p_{ln}^{\phi_{ln}}$ is the power received at LP $l$ from AP $n$ using best sector ID $\phi_{ln}$. When $\phi_{ln}$ is equal to *null*, $p_{ln}^{\phi_{ln}}$ becomes to 0. Figures 16 and 17 show examples of Wi-Fi RSS fingerprint radio maps using Wi-Fi APs located at X = 10 m, Y = 0.5 m and Z = 3 m and X = 0.5 m, Y = 0.5 m and Z = 3 m respectively with uniformly distributed LPs in a room area of 72 m². Also, Fig 18 shows the mmw radio map for a mmw AP located at X = 6 m, Y = 3 m and Z = 3. The color bars in Figs 16 and 17 indicate Wi-Fi RSS values in dBm, and in Fig 18, it indicates the mmw best sector ID, respectively. Each square in the radio maps indicates a different LP location. As it is clearly shown, both Wi-Fi RSS fingerprints and mmw best sector ID are location dependent. Consequently, the best sector ID that can cover a LP from the mmw AP can be easily indexed by the Wi-Fi RSS readings from that LP, which is the main purpose of the offline statistical phase. The accuracy of identifying the best beam ID is highly increased by increasing the number of Wi-Fi RSS readings from the LPs by using many Wi-Fi APs.

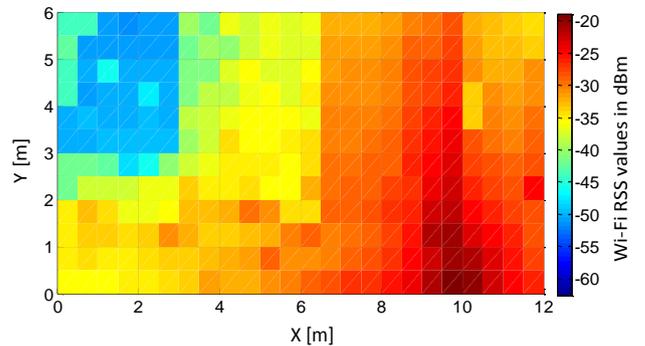

**Fig. 16** Wi-Fi RSS fingerprint radio map for AP at X = 10 m, Y = 0.5 m and Z = 3 m.

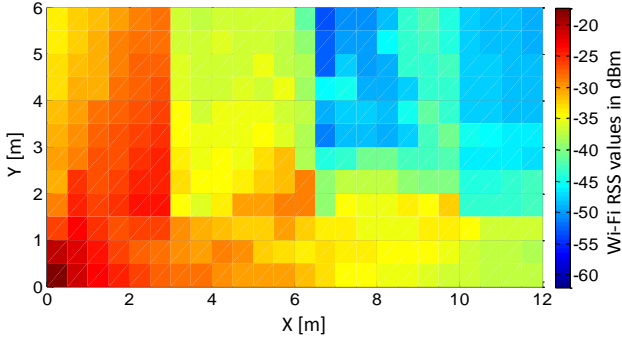

**Fig. 17** Wi-Fi RSS fingerprint radio map for AP at X = 0.5 m, Y = 0.5 m and Z=3 m.

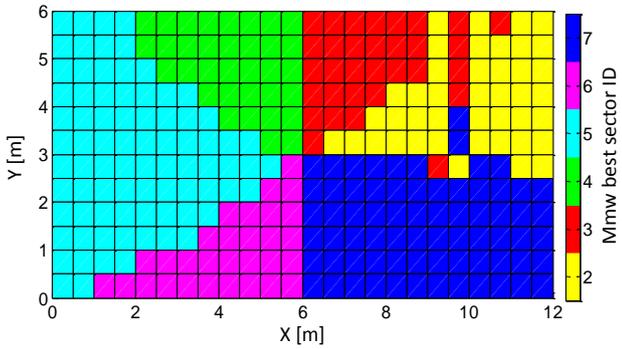

**Fig. 18** Mmw best sector ID radio map for AP at X = 6 m, Y = 3 m and Z=3 m.

### 3.3.1.2 Grouping and Clustering

As shown in Figs 16, 17 and 18, a group of LPs may be covered by the same best sector ID while they have different Wi-Fi RSS fingerprint values. Therefore, to effectively localize a best sector ID $d_n^*$ of AP $n$ and reduce the computational complexity of Wi-Fi fingerprint matching in the online phase, grouping and clustering are applied on the Wi-Fi fingerprints. Namely, the Wi-Fi fingerprints of LPs corresponding to the same best sector ID $d_n^*$ are grouped together then clustered. Because the mmw radio map given in Fig. 18 is irregular due to mmw reflections and diffractions, affinity propagation clustering algorithm [43] is used as an appropriate clustering algorithm to form the clusters. At the end of the clustering operation, we obtain a set of Wi-Fi fingerprint exemplars $\mathfrak{I}_j^{d_n^*}$, $j \in \{1,2,\dots,C_{d_n^*}\}$ for the best sector ID $d_n^*$, where $C_{d_n^*}$ is the total number of Wi-Fi fingerprint exemplars (clusters) for $d_n^*$.

### 3.3.2 Online Best Beam Selection and Bad Beams Elimination Phase

The actual mmw coordination transmissions take place in this real-time phase. During this phase, the online Wi-Fi RSS fingerprint vector $\boldsymbol{\psi}_r$ from the target UE, at an

arbitrary position $r$, is measured by the APs and collected by the APC. $\boldsymbol{\psi}_r$ is defined as:

$$\boldsymbol{\psi}_r = [\psi_{r1}\,\psi_{r2}\,\dots.\psi_{rN}]^T. \qquad (4)$$

Using $\boldsymbol{\psi}_r$, $\boldsymbol{\psi}$, $\boldsymbol{\Phi}$ and $\mathbf{P_{OFF}}$, the APC can associate the UE to an appropriate unused AP $n$. After that, the APC estimates a group of best beams for AP $n$ to communicate with the UE at its current position $r$. This can be done by calculating Euclidian distance between $\boldsymbol{\psi}_r$ and the Wi-Fi fingerprint exemplars of each best sector ID $d_n^*$. Therefore, a vector of Euclidian distance is obtained with a length up to the total number of best sector IDs. Then, the APC sorts the obtained vector of Euclidian distance in an ascending order, and it selects a group of best sector IDs (best beams) $d_n^*(1:X)$ up to $X$, where $X$ is the total number of estimated best beams, as:

$$d_n^*(1:X) = \underset{d_n^*}{\text{sort}}\left(\arg\min_{1\le j\le C_{d_n^*}}\left\|\boldsymbol{\psi}_r - \mathfrak{I}_j^{d_n^*}\right\|^2\right)_{1:X}. \quad (5)$$

After estimating the best beams $d_n^*(1:X)$, the APC pre-estimates bad beam candidates from other APs $m$ $(m \neq n)$ that may collide with these estimated best beams. In evaluating bad beam candidates, we used the following criterion:

$$d_{bm}^{d_n^*(x)}(i) = \forall_{d_m^*,m\neq n}\left(MCS_{d_n^*(x)}^{d_m^*} < MCS_{d_n^*(x)}\right),$$

$$1 \le x \le X, \ 1 \le m \le N, \ 1 \le i \le Cand_{bm}^{d_n^*(x)} \qquad (6)$$

where $MCS_{d_n^*}(x)$ is the ideal MCS value if AP $n$ uses best beam $d_n^*(x)$ to communicate with the UE without interference. $MCS_{d_n^*(x)}^{d_m^*}$ is the realized MCS value when AP $n$ uses best beam $d_n^*(x)$ while AP $m$ uses best beam $d_m^*$ at the same time. $d_{bm}^{d_n^*(x)}(i)$ is the ID of the bad beam candidate at AP $m$ that degrades the MCS of $d_n^*(x)$, and $Cand_{bm}^{d_n^*(x)}$ is the total number of bad beam candidates against $d_n^*(x)$ at AP $m$. The APC calculates (6) based on $SNR_{d_n^*(x)}$ and $SINR_{d_n^*(x)}^{d_m^*}$, where $SNR_{d_n^*}(x)$ is the received signal to noise ratio (SNR) if AP $n$ uses best beam $d_n^*(x)$, $SINR_{d_n^*(x)}^{d_m^*}$ is the SINR if $d_m^*$ interferes with $d_n^*(x)$. The APC calculates $SNR_{d_n^*}(x)$ and $SINR_{d_n^*(x)}^{d_m^*}$ for all LPs $z$ covered by AP $n$ and AP $m$ using $d_n^*(x)$ and $d_m^*$ in the $\boldsymbol{\Phi}$ matrix, which are called overlapped LPs. The calculation is done based on the $\mathbf{P_{OFF}}$ as follows:

$$SNR_{d_n^*(x)}(z) = \frac{p_{nz}^{d_n^*(x)}}{\sigma^2}, \qquad (7)$$

$$SINR_{d_n^*(x)}^{d_m^*}(z) = \frac{P_{zn}^{d_n^*(x)}}{P_{mz}^{d_m^*} + \sigma^2}, 1 \le z \le Z, \qquad (8)$$

where $p_{nz}^{d_n^*(x)}$ and $P_{mz}^{d_m^*}$ are the offline power received at overlapped LP $z$ from AP $n$ using best beam $d_n^*(x)$ and from AP $m$ using best beam $d_m^*$ respectively, $\sigma^2$ is the noise power, and $Z$ is the total number of overlapped LPs. If at least one of the overlapped LPs satisfies (6), the ID number of $d_m^*$ is registered as a bad beam candidate from AP $m$ to $d_n^*(x)$, i.e. $d_{bm}^{d_n^*(x)}(i)$. After estimating $d_n^*(1:X)$ and $d_{bm}^{d_n^*(x)}$, the APC sends $d_n^*(1:X)$ to AP $n$, and $d_n^*(1:X)$ and $d_{bm}^{d_n^*(x)}$ to the AP $m$. After NAV set using Wi-Fi, AP $n$ conducts BF refinement to find out the best beam $d_n^*(x^*)$ among $d_n^*(1:X)$, then it broadcasts its ID using the BID frame using Wi-Fi. By knowing the actual $d_n^*(x^*)$, other APs can find out more accurate bad beams $d_{bm}^{d_n^*(x^*)}(i)$ using above criterion. If AP $m$ is selected for associating another UE, it will eliminate the estimated bad beams $d_{bm}^{d_n^*(x^*)}$ from its estimated best beams $d_m^*(1:X)$ before starting BF refinement. Thus, a collision free coordinated BF training can be achieved.

### 3.4 Simulation Analysis

#### 3.4.1 Simulation Area and Simulation Parameters

In this section, the efficiency of the proposed sub-cloud WLAN compared to the conventional un-coordinated mmw WLAN is verified via computer simulations. Figure 19 shows ray tracing simulation area of a target environment, in which 8 dual band (Wi-Fi/mmw) APs are deployed on the ceiling. Simulation parameters are given in Table 4. The steering antenna model defined in IEEE 802.11ad [19] is used as the transmit antenna directivity $G(\varphi, \theta)$ for mmw APs. In which, the beam gain is designed as 25dBi with half power beamwidth of $30°$ both in azimuth $\varphi$ and elevation $\theta$ directions. Using $G(\varphi, \theta)$, the total channel gain from mmw AP in the mmw band becomes:

$$g(\tau) = \int_0^{2\pi} \int_0^{\pi} \sqrt{G(\varphi, \theta)} \, h(\varphi, \theta, \tau) \sin\theta \mathrm{d}\theta \mathrm{d}\varphi, \qquad (9)$$

where $h(\varphi, \theta, \tau)$ is the channel response calculated by the ray tracing simulation without BF gain.

In the simulation, the following two metrics are evaluated. The first one is total system throughput, which is defined as the sum of throughputs for all successfully delivered packets from all mmw APs. The second one is the average packet delay, which is defined as average time period from the instant when a packet occurs in a mmw AP to the instant when the target UE completes receiving it.

#### 3.4.2 Simulation Results

Figures 20, and 21 show average total system throughput in Gbps and average packet delay in msec, respectively, as a function of the used number of mmw APs. As it is clearly shown, the total system throughput achieved by the proposed coordinated mmw WLAN is always higher than that achieved by the conventional un-coordinated one. Also, as the number of mmw APs is increased, both performance metrics of the un-coordinated WLAN in Fig. 20 and Fig. 21

slightly degrade. This comes from the random access in the un-coordinated exhaustive search BF, which causes a lot of packet collisions as previously explained. On the other hand, using the proposed coordinated mmw transmissions, the system throughput is highly increased as we increase the number of APs. Using 8 mmw APs, about 6 times increase in total system throughput is achieved by the proposed coordinated WLAN over the un-coordinated one. Likewise, the proposed coordinated mmw WLAN greatly enhances the average packet delay. Also, using 8 APs, the average packet delay of the un-coordinated WLAN is 2.5 times longer than that of the proposed coordinated WLAN. This packet delay reduction comes from not only avoiding packet collisions but also eliminating the exhaustive search BF mechanism using statistical learning.

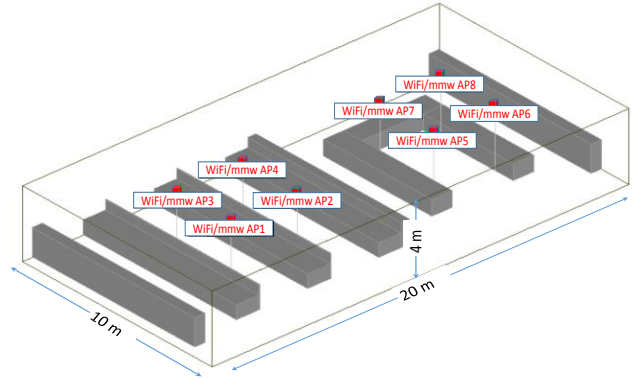

**Fig. 19** Ray tracing simulation area.

**Table 4** Simulation parameters.

| Parameter | Value |
|---|---|
| Transmission mode | SC/FDE |
| Number of mmw APs/UEs | 8/24 |
| Tx power of 5 GHz/60 GHz | 20dBm/10dBm |
| Beamwidth in azimuth and elevation | $30°, 30°$ |
| Traffic model | Poisson distribution |
| Offered load /Packet size for UE | 1Gbps /1500 octet |
| Max. num. of re-transmission | 10 |
| Num. of estimated best beams | 6 |
| Num. of antenna sectors | 36 |
| Num. of LPs | 90 |
| Beacon interval time | 1sec |
| Beam gain | 25 dBi |

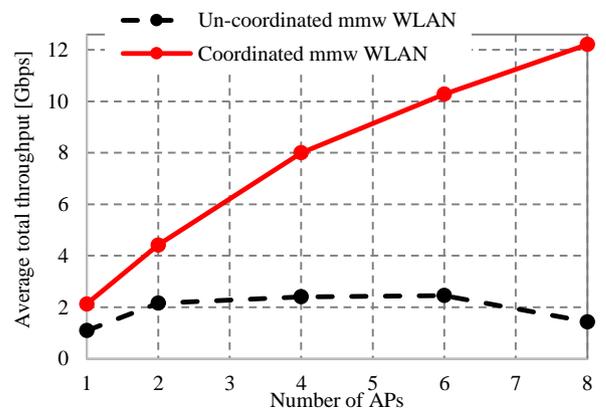

**Fig. 20** Average total throughput.

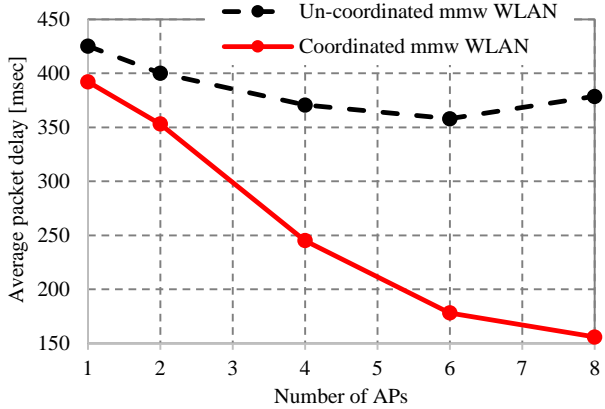

**Fig. 21** Average packet delay.

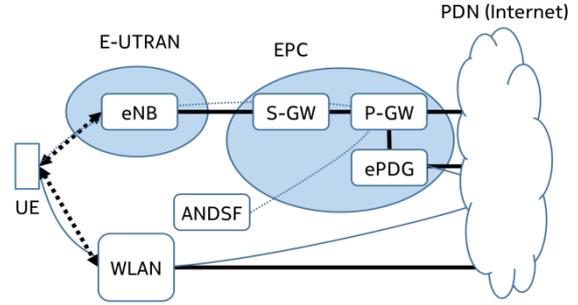

**Fig. 22** 3GPP LTE/WLAN internetworking architecture.

## 4. LTE/WiGig Internetworking for 5G cellular Networks

Although interworking between high capacity WLANs and LTE networks is a key enabler of future 5G cellular networks, there are several challenges to efficiently internetwork them. These challenges include WLAN discovery, coordination over WLANs, and financial issues in terms of capital expenditures (CAPEX) and operating expense (OPEX). The HetNet architecture and C/U splitting including the function of ANDSF are proposed to facilitate the LTE and WLAN internetworking [32]-[37]. By using these techniques, the wide coverage LTE is used to perform the signaling and a combination of LTE and WLAN opportunities is used to deliver the data. In this section, we introduce three different architectures for LTE/WLAN internetworking namely the loosely coupled in the current 3GPP standard [36], the tightly coupled, and the hybrid coupled LTE/WLAN internetworking proposed by the authors in [38] and [39].

### 4.1 Loosely Coupled 3GPP LTE/WLAN Internetworking

The main objective of the current 3GPP architecture [36], shown in Fig. 22, is to actively facilitate user data offloading from LTE to WLANs, by which traffic congestion in LTE could be relaxed. Due to this simple objective, the architecture is very simple and mainly depends on three independent parts: WLANs, ANDSF and evolved packet data gateway (ePDG), as shown in Fig. 22. WLANs are independently and directly connected to the Internet. This means that WLANs can be installed anywhere and owned by anybody, and therefore we call these WLANs as third-party radio access networks (3RANs). In order to efficiently operate 3RANs in LTE areas, the LTE uses ANDSF to partially control the offloading process through the WLANs. ANDSF is used to support users to efficiently discover and select 3RANs autonomously by giving a list of nearby WLANs over the LTE link as a very primitive version of C/U splitting [37]. ePDG works as a gateway for users to establish end-to-end secure connections to packet data network (PDN), e.g. Internet, through evolved packet core (EPC) [44]-[46]. Although this HetNet architecture has several advantages, such as simplicity and good cost performance, it lacks tight coordination over WLANs. Especially in the case of mmw WLANs, since the coverage of WLANs are very small, it is difficult to coordinate them via loosely coupled architecture due to the latency of the signalling. Energy efficient WLAN discovery protocols also cannot be achieved because users themselves scan through the list of nearby WLANs to finally decide to which ones they associate. Therefore, optimal WLAN selection cannot be done due to the lack of centralized control. As well, schemes based on data pre-fetching and delayed offloading cannot be handled [47] [48].

### 4.2 Tightly Coupled LTE/WLAN Internetworking

Figure 23 depicts the tightly coupled LTE/mmw WLAN architecture proposed by the authors in [38]. In this architecture, one evolved node B (eNB), namely LTE base station (BS), and multiple mmw WLANs are directly connected and are in master-slave relationships in a unified evolved universal terrestrial RAN (E-UTRAN) architecture. Therefore, concrete C/U splitting can be easily operated and managed using this architecture. Hence, radio resources and WLANs context information can be uniformly managed and conveyed by the eNB. Control signals are transmitted solely by the eNB, while user data could be opportunistically served by either the eNB or WLANs. In order to realize seamless LTE/mmw handover via C/U splitting, the architecture employed two LTE/WLAN protocol adaptors in mmw WLAN and UE. The function of the adaptors is to mutually translate and encapsulate IP-conformant LTE/WLAN signalling. Mmw WLANs have one half of the protocol adaptors so that the eNB and mmw WLANs are able to communicate using X2 reference point defined by 3GPP [49]. Another half of the protocol adaptors are installed on UEs, by which UEs are able to handle mmw controlling signals conveyed by the eNB and user data transmitted by mmw WLANs using upper-layer processors for mmw WLAN and LTE, respectively.

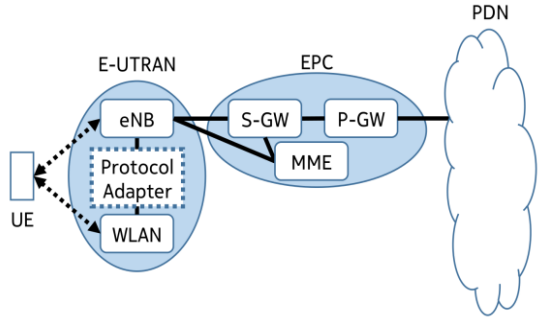

**Fig. 23** Tightly LTE/WLAN internetworking architecture.

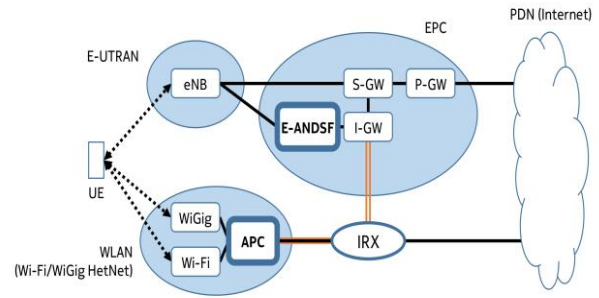

**Fig. 24** Hybrid LTE/WLAN internetworking architecture.

The strength of this architecture is that the mmw WLANs are able to be members of uniformed cloud RANs (C-RANs) [32] [33]. Uniformed C-RANs may provide uniformed user experience regardless of radio access technologies (RATs) to be used, e.g. seamless handover from LTE to WLANs and vice versa. One of the potential challenges, on the other hand, is a costing issue for RAN operation. This is because all of the BSs, the WLANs and internetworking facilities have to be constructed and maintained solely by a single mobile network operator (MNO) – this challenge becomes significant especially if the number of WLANs increases. Furthermore, standardization of control signalling between the eNB and mmw WLANs is known to be a future work.

### 4.3 Hybrid Coupled LTE/WLAN Internetworking

The proposed architecture in [39] and shown in Fig. 24 aims to stimulate 3RAN architecture to reduce the CAPEX by employing loosely coupled architecture between LTE and WLAN APCs, while APCs realize tightly coupled coordination over mmw WLANs by using sub-cloud architecture as explained in Sect. 3. The OPEX of the networks can also be reduced by sharing the 3RANs among multiple MNOs via inter-RAN exchange (IRX). This hybrid coupled architecture mainly consists of EPC, 3RANs including APCs, IRX and enhanced-ANDSF (E-ANDSF). IRX is introduced as an entity to facilitate sharing of 3RANs among different EPCs of different MNOs over virtual internetworking. To maintain the entire network confidentiality, all the partners (3RANs and EPCs operators) in IRX are required to disclose and maintain their network operation policy (NOP), especially when they coordinate a new link and exchange keys to establish secure interconnections. In this architecture, the sub-cloud WLANs discussed in Sec. 3 should be used as 3RANs to coordinate tightly over WiGig APs via Wi-Fi and APC as shown in Fig. 24. Thus, this architecture can be considered as a multi-layer LTE/Wi-Fi/WiGig HetNet topology. Cooperation in this multi-layer HetNet is monitored and controlled by E-ANDSF in each EPC and mmw WLAN APC. Many WLAN, MNO and user context information can be conveyed using E-ANDSF for concrete control over the suggested HetNet architecture. As a result, WLAN discovery and WLAN optimal selection can be achieved in a tree type centralized manner. Furthermore, data prefetching, cache control, user trajectory expectation, delayed offloading and cell broadcasting can be easily offered.

To regulate 3RAN usages among the MNOs, the authors proposed to lease and share resources at 3RANs in traffic basis. Thus, MNOs pay for the amount of actually served traffic or reserved traffic capacity and upper-stream traffic flows are regulated as contracted. This regulation model enables MNOs to moderately satisfy high quality services and high user experience, compared to the architecture given in Sect. 4.1, with much lower cost than that in Sect. 4.2, thanks to IRX and E-ANDSF. More reliable network credibility and protocols for inter-RAN cooperation in terms of authentication, authorization, and accounting could be important topics for further investigations.

## 5. Conclusion

In this paper, we focused on heterogeneous networks using millimeter wave technology as a key technology for future 5G cellular networks. Towards that, we explored the latest developments in millimeter fabrication and access point prototyping. Then, we explored the challenges of extending millimeter wave technology for WLAN applications. Due to the short range transmissions and high susceptibility to path blocking, multiple millimeter wave APs are required to cover a target environment for high capacity multi-Gbps WLANs. Beamforming coordination is one of the most challenging issues in extending millimeter wave technology for WLAN applications. To efficiently solve this problem, in this paper, we presented two coordinated millimeter wave WLAN architectures. One is a distributed antenna type architecture to realize centralized coordination, while the other is an autonomous coordination by using the assistance of legacy Wi-Fi signaling. In addition, we suggested efficient heterogeneous network architectures to embed the coordinated millimeter wave WLANs in cellular networks as a 5G extension. We believe that these works will contribute to the realization of 5G systems, especially in 2020 Tokyo Olympic.

## Acknowledgments

This work was partly supported by "The research and development project for expansion of radio spectrum resources" of The Ministry of Internal Affairs and Communications, Japan. Also, Part of this work has been done under a project named "Millimeter-Wave Evolution for Backhaul and Access (MiWEBA)" under international cooperation program of ICT-2013 EU-Japan supported by FP7 in EU and MIC in Japan.

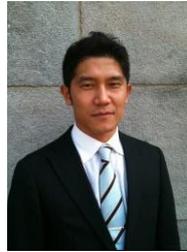

**Kei Sakaguchi** received the B.E. degree in electrical and computer engineering from Nagoya Institute of Technology, Japan in 1996, and the M.E. degree in information processing from Tokyo Institute of Technology, Japan in 1998, and the Ph.D. degree in electrical and electronic engineering from Tokyo Institute of Technology in 2006. From 2000 to 2007, he was an Assistant Professor at Tokyo Institute of Technology. Since 2007, he has been an Associate Professor at the same university. Since 2012, he has also joined in Osaka University as an Associate Professor. He received the Young Engineering Awards from IEICE and IEEE AP-S Japan Chapter in 2001 and 2002 respectively, the Outstanding Paper Awards from SDR Forum and IEICE in 2004 and 2005, respectively, the Tutorial Paper Award from IEICE Communication Society in 2006, and the Best Paper Awards from IEICE Communication Society in 2012, 2013, and 2015. He is currently playing roles of the General Chair in IEEE WDN-CN2015, the Honorary General Chair in IEEE CSCN2015, and the TPC Chair in IEEE RFID-TA2015. His current research interests are 5G cellular networks, sensor networks, and wireless energy transmission. He is a member of IEEE.

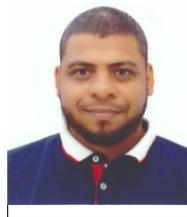

**Ehab Mahmoud Mohamed** received the B.E. degree in electrical engineering from South Valley University, Egypt, in 2001, and the M.E. degree in computer science from South Valley University, Egypt in 2006, and the Ph.D. degree in Information Science and electrical engineering from Kyushu University, Japan in 2012. From 2003 to 2008, he was an Assistant Lecturer at South Valley University, Egypt. Since 2012, he has been an Assistant Professor at Aswan University, Egypt. Since 2013, he has also joined in Osaka University, Japan as a Specially Appointed Researcher. He is a technical committee member in many international conferences and a reviewer in many international conferences and transactions. His current research interests are 5G networks, cognitive radio networks, millimeter wave transmissions and MIMO systems. He is an IEEE member.

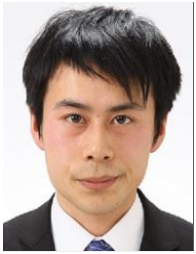

**Hideyuki Kusano** was born in 1989. He received the B.E. degree in electrical and electronic engineering from the Tokyo Institute of Technology, Tokyo, Japan, in 2013, and the M.E. degree in electrical, electronic and information engineering from the Osaka University, Osaka, Japan, in 2015, where he conducted the research of next generation millimeter-wave WLAN. In 2015, he joined the NEC Corporation, Tokyo, Japan. Currently, he is engaged in new business development, especially in IoT/M2M market.

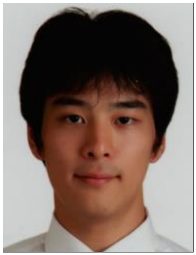

**Makoto Mizukami** was born in 1990. He received a B.E. degree in electronic and information engineering from Osaka University in 2014. He had been a research student at Tokyo Institute of Technology until 2015, where he conducted researches on cellular networks and software radio. In 2015, he has started his studies in user interface and human cognition, expecting to receive a M.E. degree in bioinformatic engineering from Osaka University in 2017. He is a student member of IEEE and IEICE.

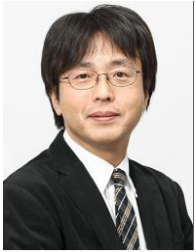

**Shinichi Miyamoto** received the B.E., M.E. and Ph.D. degrees in communications engineering from Osaka University, Japan in 1990, 1992 and 1998, respectively. From 1993-2005, he was an assistant professor, and from 2005-2015 he was an associate professor at the Graduate School of Engineering, Osaka University. In 2015, he joined Wakayama University as a professor at the faculty of systems engineering. During 2001-2002, he was at the Virginia Institute of Technology as a Visiting Professor. He received the Young Researchers' Award from the IEICE of Japan in 1998. He is a member of IEEE and ITE.

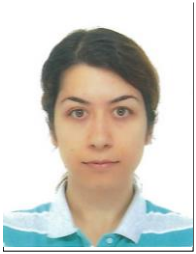

**Roya Rezagah** received B.Sc. in Electrical Engineering from Sharif University of Technology, Tehran, Iran in 2002, and M.Sc. and Ph.D. in Communications systems from Amirkabir University of Technology (Tehran Polytechnic), Tehran, Iran, in 2006 and 2011 respectively. During her study in Tehran Polytechnic, she joined several industrial projects as a researcher and system designer. Those industrial experiences cover a wide range from RFID to cell planning and DVB-T implementation. In 2011, she joined Araki-Sakaguchi lab. in the Department of Electrical and Electronic Engineering, Tokyo Institute of Technology, Tokyo, Japan and on September 2014, she received D.Eng. from Tokyo Institute of Technology. Since 2011, she has been working on various issues of radio resource management for 5G multiband cellular networks in Araki-Sakaguchi lab.

with close collaboration with industry and mobile service providers.

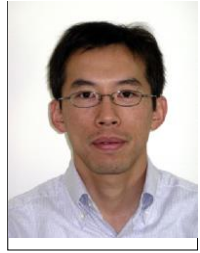

**Koji Takinami** received the B.S. and M.S. degrees in electrical engineering from Kyoto University, Kyoto, Japan, in 1995 and 1997, respectively, and the Ph.D. degree in physical electronics from Tokyo Institute of Technology, Tokyo, Japan, in 2013. In 1997, he joined Matsushita Electric Industrial (Panasonic) Co., Ltd., Osaka, Japan. Since then he has been engaged in the design of analog and RF circuits for wireless communications. From 2004 to 2006, he was a visiting scholar at the University of California, Los Angeles (UCLA), where he was involved in the architecture and circuit design of the high efficiency CMOS power amplifier. In 2006, he joined Panasonic Silicon Valley Lab, Cupertino, CA, USA, where he worked on high efficiency transmitters and low phase-noise digital PLLs. In 2010, he relocated to Japan and currently leads the development of the millimeter wave transceiver ICs. Dr. Takinami is a co-recipient of the Best Paper Award at the 2012 Asia-Pacific Microwave Conference. He was a member of the IEEE ISSCC Technical Program Committee from 2012 to 2014.

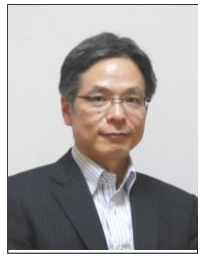

**Kazuaki Takahashi** received the B.E. and M.E. degrees in electrical and computer engineering, and the Ph.D. degree in electrical engineering from Yokohama National University, Yokohama, Japan, in 1986, 1988 and 2006, respectively. In 1988, he joined the Tokyo Research Laboratory, Matsushita Electric Industrial Co. Ltd., Kawasaki Japan, where he was engaged in research and development of monolithic microwave ICs, millimeter wave ICs based on Si and GaAs for mobile communication equipment. His current research interests include the development of short range multi-gigabit wireless system and high resolution radar system in millimeter wave and terahertz bands, and low-power radio systems for IoT/M2M. He is also working on developing wireless standards in IEEE and Wi-Fi Alliance. He is currently engaged in Platform Technology Development Center, Automotive and Industrial Systems Company, Panasonic Corporation, Yokohama Japan. Dr. Takahashi is a member of the IEEE.

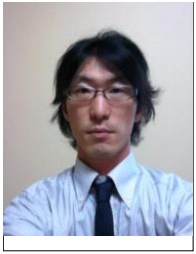

**Naganori Shirakata** received B.E and M.E. degrees in electronics engineering from Kyoto Institute of Technology, Japan in 1993 and 1995 respectively. He joined Panasonic Corporation, (former Matsushita Electric Industrial Co., Ltd.), Osaka in 1995, where he worked on baseband signal processing of OFDM and MIMO for WLANs. From 2008 to 2010, He worked on ultra low power radio and WBAN systems. From 2010 to 2013, he worked on the development of the PHY system and circuit design for millimeter wave transceiver. He currently leads the development of the system design for millimeter wave access point and network. He is a member of IEEE.

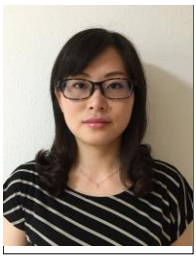

**Hailan Peng** received her B.E. and M.E. degrees in Electrical Engineering from Beijing University of Posts and Telecommunications, Beijing, China, in 2006 and 2009, respectively, and Ph.D. degree in Electrical Engineering from The University of Electro-Communications, Tokyo, Japan, in 2013. Since 2013, she joined KDDI R&D Laboratories as a research engineer. Her current research interests include 5G cellular networks, millimeter-wave communications, Heterogeneous networks and 3GPP/non-3GPP interworking and aggregation. She also contributes to 3GPP standardization activities and serves as technical program committee reviewer for IEICE journal. She is a member of IEICE and IEEE.

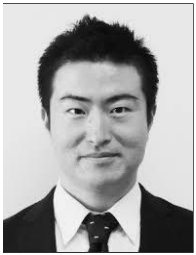

**Toshiaki Yamamoto** received the B.E. degree in Electrical and Electronic Engineering, M.I. and Ph.D. degrees in Informatics from Kyoto University, Kyoto, Japan in 1999, 2001, and 2004, respectively. In 2004, he joined KDDI R&D Labs. and has been engaged in research on wireless communication technologies for LTE and LTE-Advanced systems. He is now assistant manager in KDDI corporation. Dr. Yamamoto received the Young Researcher's Award of IEICE Japan in 2008. He is a member of the IEEE.

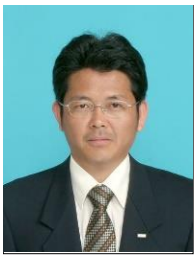

**Shinobu Namba** received the B.E., M.E., and Ph.D. degrees, in information science and electrical engineering from Kyushu University, Fukuoka, Japan, in 1994, 1996, and 2006, respectively. He joined Kokusai Denshin Denwa Co., Ltd. (now KDDI Corp.) in 1996. He is currently a R&D manager of Optical Access Network Laboratory in KDDI R&D Laboratories, Inc. He received the Young Researcher's Award from the Institute of Electronics Information, and Communication Engineers (IEICE) in 2002, a best paper award of the 16th Asia-Pacific Conference on Communications (APCC 2010) in 2010 and a best paper award of the 5th European Conference on Antennas and

Propagation (EuCAP 2011) in 2011.